# Intense terahertz radiation via the transverse thermoelectric effect

*Petar Yordanov[1*], Tim Priessnitz[1,2*], Min-Jae Kim[1,2,3], Georg Cristiani[1], Gennady Logvenov[1], Bernhard Keimer[1], and Stefan Kaiser[1,2,3]*

(1) Max-Planck Institute for Solid State Research, 70569 Stuttgart, Germany

(2) 4-th Physics Institute and Research Center SCoPE, University of Stuttgart, 70569 Stuttgart, Germany

(3) Technical University Dresden, Institute of Solid State and Materials Physics, 01069 Dresden, Germany

* These authors contributed equally to this work.

**Terahertz (THz) radiation is a powerful tool with widespread applications ranging from imaging, sensing, and broadband communications to spectroscopy and nonlinear control of materials [1]. Future progress in THz technology depends on the development of efficient, structurally simple THz emitters that can be implemented in advanced miniaturized devices. Here we show how the natural electronic anisotropy of layered conducting transition metal oxides enables the generation of intense terahertz radiation via the transverse thermoelectric effect. In thin films grown on offcut substrates, femtosecond laser pulses generate ultrafast out-of-plane temperature gradients, which in turn launch in-plane thermoelectric currents, thus allowing efficient emission of the resulting THz field out of the film structure. We demonstrate this scheme in experiments on thin films of the layered metals $PdCoO_2$ and $La_{1.84}Sr_{0.16}CuO_4$, and present model calculations that elucidate the influence of the material parameters on the intensity and spectral characteristics of the emitted THz field. Due to its simplicity, the method opens up a promising avenue for the development of highly versatile THz sources and integrable emitter elements.**

Miniaturization of functional components is crucial for the realization of integrated terahertz (∼0.1 − 30 THz) technology [1]. In particular, advances in THz emitters are driven by the discovery of new materials and generation mechanisms that enable their integration into compact circuit modules. Currently, portable sources of intense THz radiation commonly rely on external femtosecond laser oscillators. The most widely used designs utilize ultrafast currents of photo-excited charge carriers in semiconductors (photoconductive antennas) or optical rectification (OR) in nonlinear crystals [2,3]. In view of the growing number of applications of THz technology, however, researchers are pursuing a diverse set of strategies to devise new concepts for THz sources with optimal efficiency, size, ease of fabrication, flexibility regarding operating conditions, and compatibility with various planar hybrid structures. Recent pertinent examples include spintronic emitters based on ultrafast photoexcitation of spin currents in ferromagnetic-nonmagnetic metallic multilayers [4,5], and quantum cascade lasers operating at elevated temperatures [6].

Here we take advantage of the intrinsic thermopower anisotropy of naturally layered materials and the properties of the transverse thermoelectric effect (TTE) [7-20] (Supplementary Text), and show that in suitably prepared thin films, small temperature differences created by femtosecond pulsed laser illumination can induce electric current transients, which give rise to emission of intense THz radiation. The method operates with monolithic thin films without the need for heteroepitaxy or extensive processing, and implies virtually no restrictions of the operating conditions.

The concept is optimally realized in layered materials with a large difference in the Seebeck coefficients between the main crystallographic axes, $\Delta S = S_{ab} - S_c$. Films of such materials are deposited on offcut substrates such that the $c$-axis is inclined at an angle $\theta$ relative to the surface normal, thus creating two inequivalent planar axes ($x$ and $y$) (Fig. 1a). Whereas transport along the $y$ axis is determined by the $ab$-plane properties of the material, transport along $x$ reflects a mixture of the $ab$-plane and $c$-axis properties which is tunable via the angle $\theta$. When a laser pulse with diameter $l$ arrives at a film with thickness $d \ll l$, it instantly heats a layer with thickness of the order of the optical penetration depth [14], thus establishing a temperature gradient between the front and the back sides of the film, $\nabla_z T = \Delta T_z/d$. By virtue of the transverse thermoelectric effect, the difference $\Delta T_z$ actuates thermal diffusion of mobile charge carriers along $x$ via a thermoelectric voltage $U_x$, which can be expressed as [7];

$$U_x = \frac{l}{d}\frac{\sin 2\theta}{2}(S_{ab} - S_c)\Delta T_z. \qquad (1)$$

The aspect ratio $l/d$ enhances $U_x$ by a factor of $\sim 10^4 - 10^5$ compared to the conventional longitudinal response $U_z$, whereas $U_y$ vanishes by symmetry. The voltage dynamics $U_x(t)$ is determined by the characteristic time for heat diffusion through the film, $\tau_r \sim d^2/D_z$, where $D_z = \kappa_z/\rho_m c_p$ is the thermal diffusivity, $\kappa_z$ the thermal conductivity, $\rho_m$ the density, and $c_p$ the specific heat of the material [10]. The corresponding time-dependent electric current density in the plane of the film, $j_x(t) \sim U_x(t)/\rho_x$ where $\rho_x$ is the effective electrical resistivity, gives rise to $z$-axis directed dipole emission in the THz domain, $E_{\text{THz},z}(t) \sim \partial j_x(t)/\partial t$ [21], if $\tau_r \lesssim 1$ ps and the laser pulses are sufficiently short ($\tau_{\text{lp}} \lesssim \tau_r$).

In light of these considerations, we have chosen the layered delafossite PdCoO$_2$ to test the TTE-driven THz generation scheme. PdCoO$_2$ has recently received considerable attention in view of its exceptionally high in-plane electrical and thermal conductivities [22-26], as well as its strongly anisotropic and ambipolar Seebeck coefficient (the key controlling parameter according to Eq. 1) predicted by ab-initio calculations [27,28].

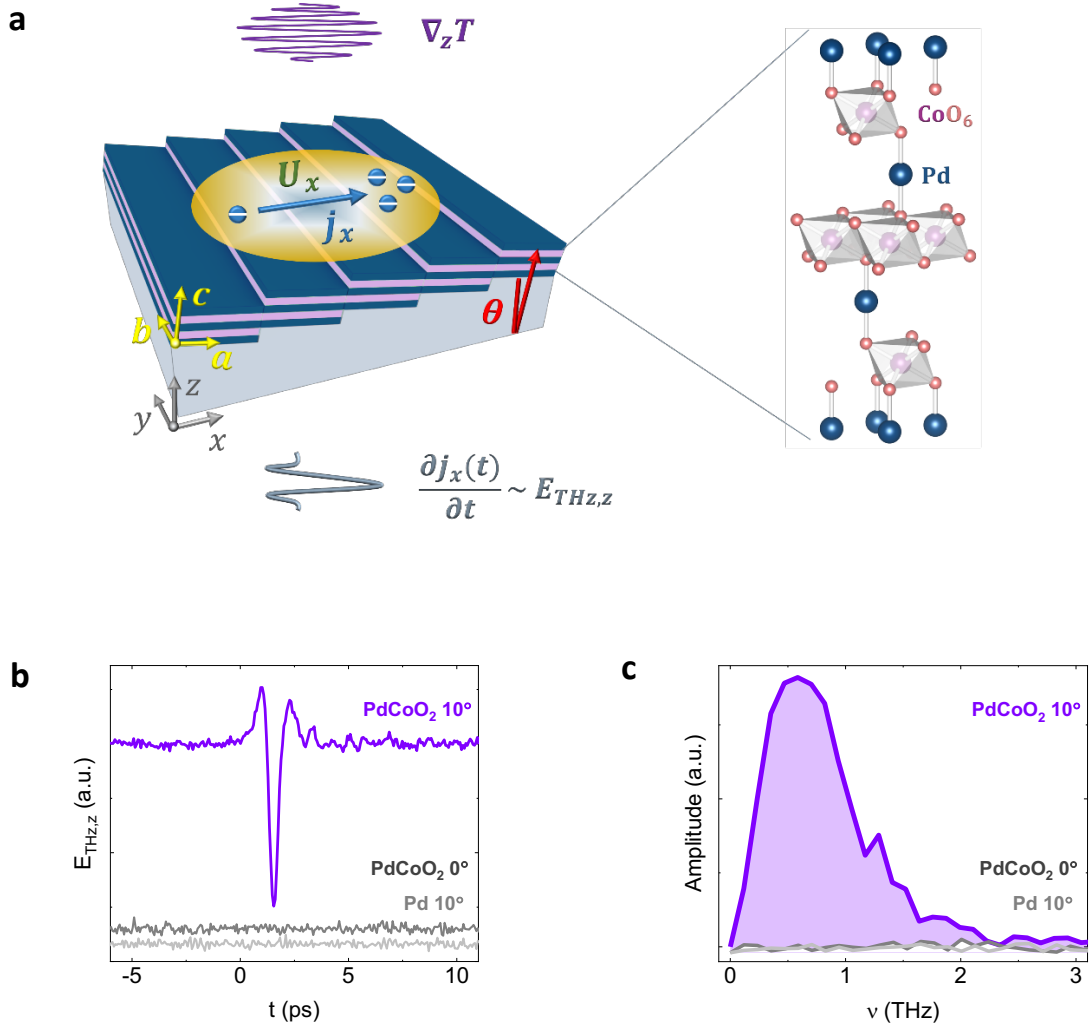

**Fig. 1: Generation of terahertz radiation via the transverse thermoelectric effect. a**, Thin film of a material with anisotropic Seebeck coefficient deposited on an offcut substrate for TTE-driven THz generation. Inset: Crystal structure of the delafossite $PdCoO_2$, comprising metallic Pd and insulating $CoO_2$ layers, with highly anisotropic structural and transport properties [22-33]. Electric field strength **(b)** and amplitude spectrum **(c)** emitted from a thin film ($d \sim 10$ nm) of $PdCoO_2$ on a $\theta = 10°$ offcut $Al_2O_3$ substrate, as described in the text. Reference measurements on a $PdCoO_2$ film grown on a regular $\theta = 0°$ $Al_2O_3$ substrate, and on an elemental Pd film on a $\theta = 10°$ $Al_2O_3$ substrate are also shown.

At room temperature, the resistivity is $\rho_{ab}/\rho_c \approx 2.6/1070$ μΩ cm [23,24], the thermal conductivity $\kappa_{ab}/\kappa_c \approx 300/50$ W $K^{-1}$ $m^{-1}$ [26], and the thermal diffusivity $D_{ab}/D_c = 1.1 \times 10^{-4}/1.8 \times 10^{-5}$ $m^2$ $s^{-1}$ [30]. Recent experiments on $PdCoO_2$ thin films and powder compacts supported the theoretical predictions for the thermopower [29-31] showing $S_{ab}/S_c \approx 4/(-30)$ μV $K^{-1}$, i.e. a difference of $\Delta S = (S_{ab} - S_c) \approx 34$ μV $K^{-1}$, at room temperature. We used pulsed laser deposition to synthesize films with $d \approx 10$ nm and $0° \leq \theta \leq 25°$ on optically transparent $Al_2O_3$ substrates [29,31,32] (Methods). Measurements of the transverse voltage $U_x(t)$ induced by ultraviolet laser pulses with $\tau_{lp} \approx 5$ ns yielded a substantial amplitude and negligible broadening of the voltage signal, thus providing direct confirmation of the large thermopower anisotropy in $PdCoO_2$ and indicating a characteristic response time $\tau_r \ll 5$ ns for the films (Supplementary Text).

To check whether the laser-induced transverse voltage elicits emission of THz radiation, we employed a standard ZnTe detector to monitor the time-dependent THz amplitude emitted from the $PdCoO_2$ films in response to optical laser pulses with $\tau_{lp} \approx 250$ fs and fluence $F_{lp} = 0.5$ mJ cm$^{-2}$ (Fig. 1a) (Methods). Figure 1b indeed shows a clear single-cycle THz electric field waveform $E_{THz,z}(t)$ emitted from a film with $\theta = 10°$. The corresponding spectrum (Fig. 1c) exhibits a broad bandwidth with a range of about $0.1 - 2.3$ THz and peak amplitude at $0.6$ THz. To quantify the THz field strength, we performed reference measurements on a commercial ZnTe OR source whose peak field at pump fluence $F_{lp} = 0.5$ mJ cm$^{-2}$ is in the range $E_{THz,z} \approx 10^5 - 10^6$ V m$^{-1}$ [34,35]. Figure 2 shows that the amplitude of radiation emitted from the $PdCoO_2$ source is only a factor of four lower than the reference source, which implies a peak field strength of $E_{THz,z} \approx 10^5$ V m$^{-1}$. The high intensity of the TTE-driven THz emission is remarkable because the $PdCoO_2$ data were taken on a monolithic, unprocessed film. This finding already indicates a high potential for optimization of the intensity, which is further discussed below.

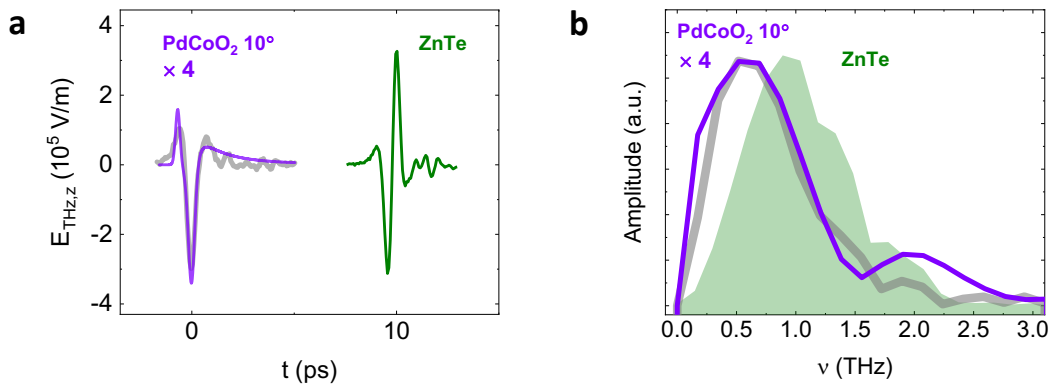

**Fig. 2: Determination of the THz electric field strength. a, b,** Comparative plots of the electric field strength and spectra of THz radiation from a $d = 10$ nm, $\theta = 10°$ $PdCoO_2$ film (gray lines: experimental data, magenta lines: model calculations described in the text) and from a ZnTe OR source (experimental data), for $\tau_{lp} \approx 250$ fs and $F_{lp} = 0.5$ mJ cm$^{-2}$.

We now present the results of a series of experiments designed to demonstrate that the observed THz radiation indeed originates from the TTE mechanism. In control experiments, we repeated the measurements on films for which this mechanism is expected to be inactive (Fig. 1b,c). Specifically, we confirmed the absence of any THz signal from a PdCoO$_2$ film grown on a regular $\theta = 0°$ substrate, as expected based on Eq. 1. This result also rules out the conventional longitudinal Seebeck effect as the origin of the observed effect [36]. In additional experiments, we found vanishing THz emission from a film of elemental Pd (which exhibits an isotropic crystal structure and no anisotropy of the Seebeck coefficient) grown on an offcut substrate with $\theta = 10°$. This finding also demonstrates that the observed THz emission from PdCoO$_2$ is unrelated to the specific step-terrace structure of the films.

In a second set of experiments (Fig. 3a,b), we measured the polarity of $E_{\text{THz},z}$ in different experimental geometries and compared the results to Eq. 1, keeping in mind that according to the TTE scenario the THz field follows the thermoelectric voltage and source current, and thus shares their polarity. Specifically, we show that $180°$ rotation of the sample around the surface normal (corresponding to $\theta \to -\theta$ in Eq. 1), reverses the polarity of $E_{\text{THz},z}$. On the other hand, changing from front- to backside illumination (which corresponds to $\Delta T_z \to -\Delta T_z$ and $\theta \to -\theta$ due to the symmetry of the tilted film structure) leaves the polarity unaffected, in agreement with Eq. 1.

In a third set of experiments, we used measurements of the light-field polarization to demonstrate two key hallmarks of the TTE mechanism. First, we determined the polarization state of the emitted THz field by inserting a polarizer in front of the detector and rotating the sample around the surface normal, as shown in Fig. 3c. The observed $90°$ periodicity of the detected intensity as a function of rotation angle (Fig. 3d) confirms that the THz field is linearly polarized along *x*, as expected from the flow of thermoelectric currents, which is confined to the *x*-axis by symmetry. Second, we confirmed that the THz signal is independent of the polarization of the pump pulse (Fig. 3e). This observation rules out surface-field or optical rectification mechanisms of THz generation [37] and demonstrates a distinguishing feature of the TTE mechanism, where the laser pulse serves exclusively as a heat source.

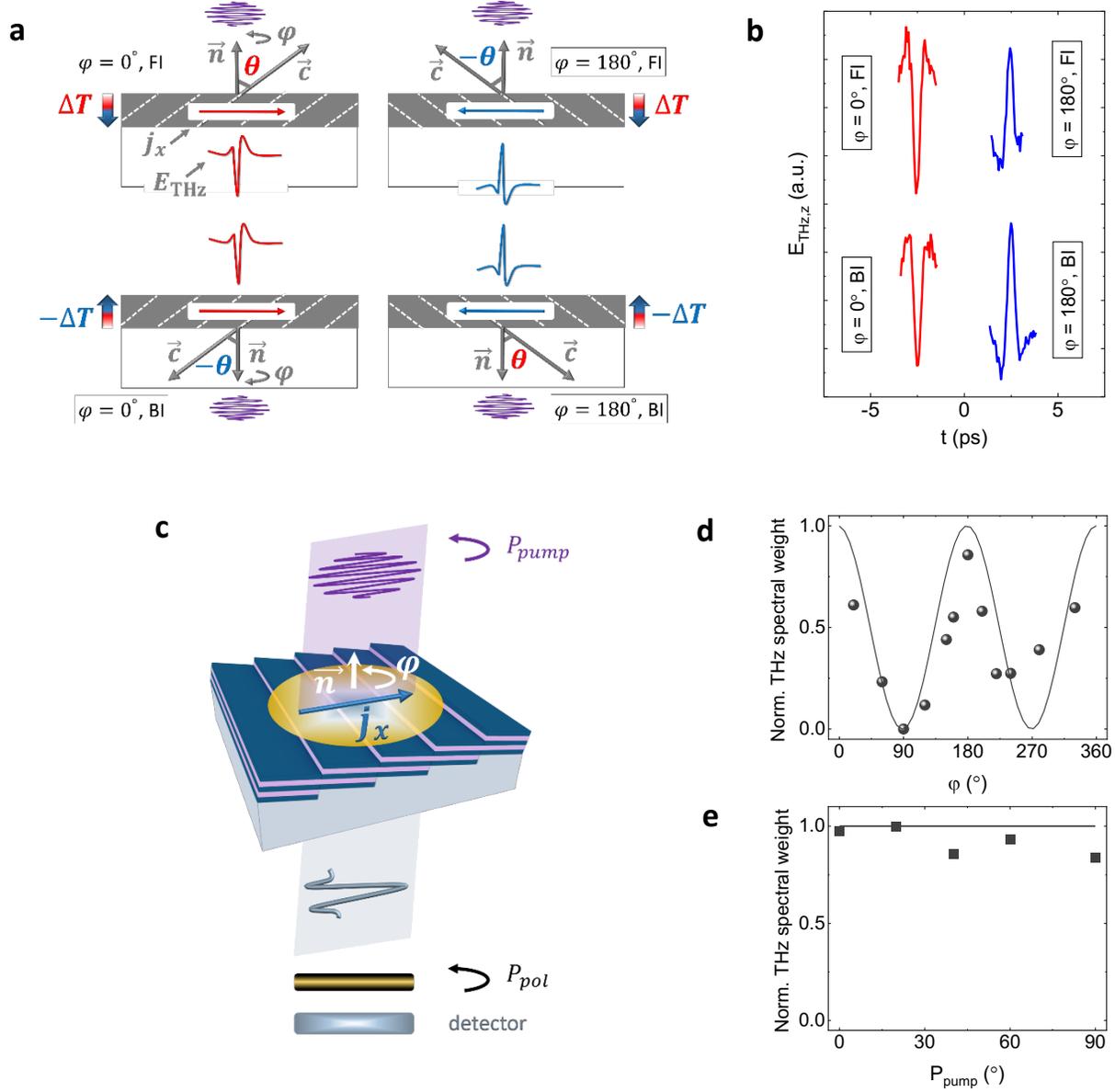

**Fig. 3: Confirmation of the TTE mechanism for THz generation. a**, Current density ($j_x$) direction and electric field ($E_{\text{THz},z}$) polarity upon $\varphi = 0° \rightarrow 180°$ rotation about the surface normal $\vec{n}$ (equivalent to $\theta \rightarrow -\theta$ in Eq. 1) (left to right), and upon change from front-side (FI) to back-side (BI) illumination (equivalent to a combined operation of both $\Delta T_z \rightarrow -\Delta T_z$ and $\theta \rightarrow -\theta$ in Eq. 1) (top to bottom). **b**, Experimental $E_{\text{THz},z}(t)$ waveforms emitted from the PdCoO$_2$ film in the four different configurations in panel a. **c**, Experimental setup to investigate the polarization state of the emitted THz field and the dependence on the pump polarization. **d**, Normalized THz spectral weight as a function of $\varphi$ for fixed pump polarization $P_{\text{pump}} = 0°$ and polarizer angle $P_{\text{pol}} = 0°$. The line denotes a sinusoidal fit of the data points. **e**, Normalized THz spectral weight as a function of the pump polarization $P_{\text{pump}}$, for fixed $\varphi = 0°$ and $P_{\text{pol}} = 0°$. The line indicates the behavior expected if the intensity does not depend on $P_{\text{pump}}$.

Having firmly established the TTE mechanism of THz generation, we now present the results of additional experiments aimed at gaining a more in-depth understanding of the factors influencing the amplitude and spectral characteristics of the emitted THz field. Figures 4a,b display data from thin films of PdCoO$_2$ deposited on $\theta = 10°, 15°$ and $25°$ offcut substrates. A clear trend of declining peak field and spectral amplitude at larger $\theta$ is visible, indicating that the increase of the thermoelectric voltage $U_x$ with $\theta$ up to $45°$ predicted by Eq. 1 is overcompensated by the $\theta$-dependences of the other effective quantities contributing to the THz source currents $j_x$ and field $E_{\text{THz},z}$ (Supplementary Text).

To gain further insight into the influence of materials parameters, we have studied TTE-induced THz generation in thin films of the layered metal La$_{1.84}$Sr$_{0.16}$CuO$_4$ (LSCO) (see Methods), whose thermopower anisotropy is comparable to that of PdCoO$_2$ ($\Delta S \sim 40$ μV K$^{-1}$) while its electrical resistivity is substantially higher ($\rho_{ab}/\rho_c \sim 150/30000$ μΩ cm) and its thermal diffusivity an order of magnitude lower ($D_{ab}/D_c \sim 3 \times 10^{-6}/1.5 \times 10^{-6}$ m$^2$ s$^{-1}$) [17, 38-40]. Figure 4 shows that the spectrum emitted from an LSCO film with $d \sim 10$ nm and $\theta = 15°$ is qualitatively similar to the corresponding PdCoO$_2$ data, which suggests that the TTE mechanism is common to layered metal oxides. The different materials parameters are reflected in a reduced THz field strength and enhanced spectral range compared to PdCoO$_2$, highlighting the potential of materials exploration in the development of functional devices.

Guided by these observations, we have devised a model for the THz far field, $E_{\text{THz},z}(t) \sim [1/(1+n_z)] \times \partial j_x(t)/\partial t$ [21] where $n_z$ refers to the refractive index at THz frequencies, in response to the thermoelectric voltage calculated according to Eq. 1, with the full set of anisotropic materials parameters for both PdCoO$_2$ and LSCO (Supplementary Text). Figure 4 shows that the model calculations yield excellent descriptions of the experimental data in both time and frequency domains. Deviations between calculated and measured data for the PdCoO$_2$ film on a $\theta = 25°$ substrate are likely a consequence of structural defects that degrade the electronic properties at high offcut angles. The dominant mechanism which drives the decrease of the peak $E_{\text{THz},z}$ and amplitude at large $\theta$ is ascribed to the increasing contribution of the more resistive $c$-axis transport ($\rho_{ab} \ll \rho_c$) to $\rho_x$. Similarly, the reduced THz intensity observed in the LSCO film is explained by the higher overall resistivity of this material compared to that of PdCoO$_2$.

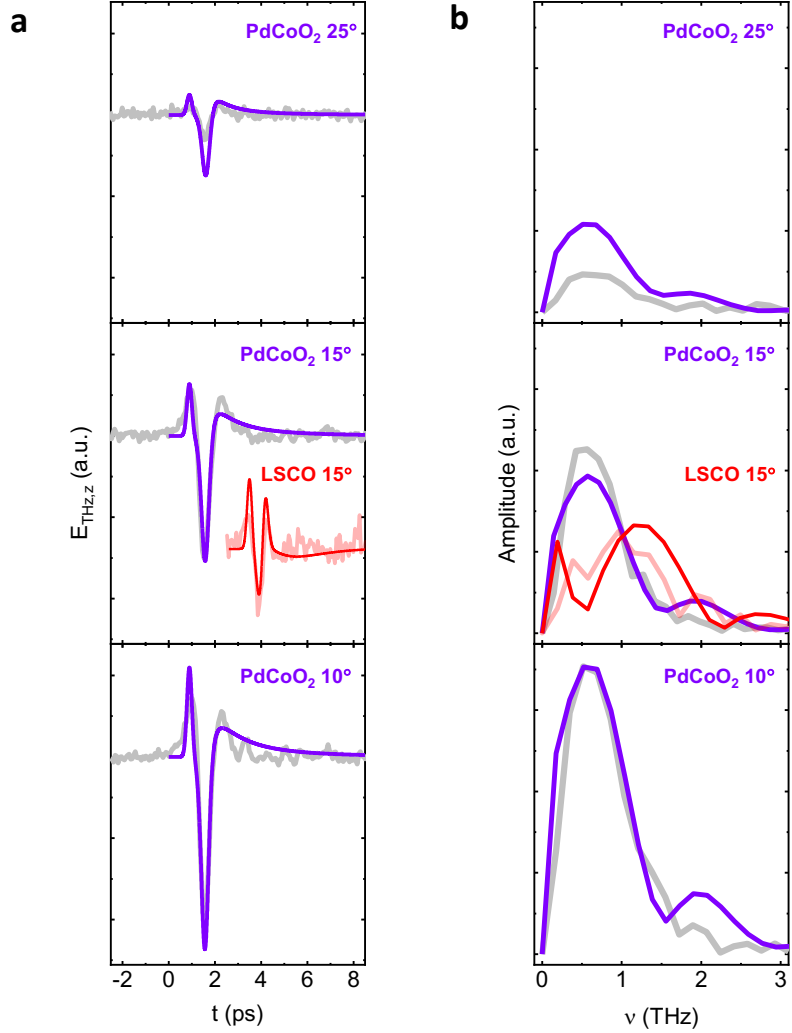

**Fig. 4: Experimental data and model calculations of TTE-THz generation in thin films of PdCoO$_2$ and La$_{1.84}$Sr$_{0.16}$CuO$_4$ - offcut angle dependence. a,b** Experimental (light grey and light red) and calculated (violet and red) THz field $E_{\text{THz},z}(t)$ and amplitude spectra from PdCoO$_2$ and La$_{1.84}$Sr$_{0.16}$CuO$_4$ (LSCO) films with $d \sim 10$ nm on substrates with different offcut angles.

An essential component in the model is the transient thermal diffusivity $D_z^*$, which depends on the laser pulse duration $\tau_{lp}$ and fluence $F_{lp}$, and has been shown to reach values three or more orders of magnitude larger than those in equilibrium [41-43]. In metallic systems, the transient thermal diffusivity is expected to thermalize through scattering with a characteristic time $\tau_{\text{th}}$ that ranges from a few tens to a few hundreds of femtoseconds [44]. These experimental findings are significant for TTE-THz generation, because they indicate that manipulation of the laser pulse characteristics can be used not only to manipulate the THz field strength, but also to enhance the thermal diffusivity and extend the THz bandwidth. In the model, we define a time-dependent function of the form $D_z(t) = D_z^0 + (D_z^* - D_z^0)e^{-\frac{t}{\tau_{\text{th}}}}$ with $D_z^*$ up to $5000 \times D_z^0$ and $\tau_{th} = 70 - $ 120 fs, which results in two-component profiles for the temperature difference $\Delta T_z(t)$ and current density $j_x(t)$. The corresponding THz waveforms thus comprise fast transient and slower equilibrium components corresponding to the spectral ranges below and above 0.5 THz, respectively. Whereas this separation of scales is clearly visible in the shape of the spectra of LSCO, the two regimes overlap in PdCoO$_2$ due to the much higher equilibrium thermal diffusivity.

The model calculations point out various perspectives for enhancement of the TTE-driven THz generation. In particular, the optimal value of the offcut angle $\theta$ is determined by a complex superposition of contributions from all relevant anisotropic transport coefficients. Based on bulk input model parameters for PdCoO$_2$, the maximum of $E_{\text{THz},z}(\theta)$ is expected in the range $3° \leq \theta \leq 7°$, which is quite comparable with the trend in the experimental data indicating a peak field at $\theta \approx 10°$ (Figs. 4 a,b and Supplementary Text). We attribute the discrepancy to differences between the anisotropy ratios of the material parameters of the bulk (which were used as inpit for the model calculations) and the thin films. We note that the thickness of the films used in our experiments ($d = 10$ nm) is well below the optical penetration depth ($\delta_{ab} \sim 70$ nm, $\delta_c \sim 100$ nm at wavelength 800 nm) [33], so that the pump intensity should be uniform across the film. Whereas the temperature difference $\Delta T_z$ under these conditions is presumably maintained by factors such as lateral heat dissipation and the asymmetry of the air-film and film-substrate interfaces, our model predicts an enhancement of $\Delta T_z$ and a corresponding enhancement of $E_{\text{THz},z}$ in the regime $d > \delta_z$. In particular, a peak field strength of $E_{\text{THz},z} \sim 1 \times 10^6$ V m$^{-1}$ and a spectral range of $0.1 - 50$ THz was calculated for PdCoO$_2$ films with $d = 100$ nm, $\theta < 10°$ and $D_z^* = 5000 \times D_z^0$ in response to pump pulses with $F_{\text{lp}} = 0.5$ mJ cm$^{-2}$ (as in our experiments) and $\tau_{\text{lp}} = 10$ fs (shorter than in the experiments). While we have thus far been unable to synthesize films of this thickness with sufficient quality, there is obviously a large potential for optimization of PdCoO$_2$ thin-film systems and for exploration of other layered materials for TTE-driven THz generation.

In conclusion, we have shown that the natural anisotropy of the transport properties of layered metal oxides enables the generation of intense THz radiation via the transverse thermoelectric effect. We demonstrated this scheme in thin films of PdCoO$_2$ and La$_{1.84}$Sr$_{0.16}$CuO$_4$, which yielded THz field strengths and spectral bandwidths already comparable to those of commercial ZnTe OR sources. The experimental scheme does not require elaborate fabrication methods and complementary microstructure elements, application of a bias voltage, an external magnetic field, or excessive laser fluence, and it is not limited to a particular material, temperature range, or specific operational conditions. The comparison of two different materials and the complementary model calculations we have reported indicate multiple tuning parameters for manipulation of the intensity and bandwidth of the THz field. TTE-THz generation may thus provide a powerful basis for the development of versatile THz sources and single-layer emitter elements. Further intriguing perspectives may emerge from possible static or dynamic interactions of the THz source current and field with various electronic states and correlations, such as high-temperature superconductivity, magnetism, and ferroelectricity, which can be explored by integration of ultrathin films of the emitter material into artificial superlattices with different transition metal oxides [45].

**Methods**

**Thin-film synthesis.** Pulsed laser deposition was used to grow $PdCoO_2$ films with thickness $d \sim 10$ nm on regular $c$-axis oriented $\theta = 0°$ (0001) and offcut $\theta = 5 - 25°$ (0001 → $11\bar{2}0$) $Al_2O_3$ substrates. The growth conditions include stoichiometric single-phase $PdCoO_2$ polycrystalline targets, oxygen pressures of $2 - 3$ mbar, a substrate temperature of $\sim 620$ °C, and laser energy density $1.9$ J cm$^{-2}$ with repetition rate $3 - 5$ Hz. $La_{1.84}Sr_{0.16}CuO_4$ films with $d \sim 10$ nm on $LaSrAlO_4$ substrates with offcut angle $\theta = 15°$ were synthesized by atomic layer-by-layer ozone assisted molecular-beam epitaxy (DCA Instruments). The substrate temperature determined by a radiation pyrometer was 650 °C, and the background ozone-oxygen pressure was $10^{-5}$ Torr. The atomic layer-by-layer growth was monitored by using in situ reflection high electron energy diffraction (RHEED). The film quality was confirmed by high-resolution X-ray diffraction, atomic force microscopy (AMF) and scanning transmission electron microscopy (STEM).

**Terahertz emission spectroscopy.** The measurements were performed by using femtosecond ($\tau_{lp} \approx 250$ fs) optical pulses (central wavelength $\lambda_{lp} = 800$ nm) - generated by a Ti:Sapphire amplifier (Coherent RegA 9000) at a repetition rate of 150 kHz and pump fluence of $F_{lp} \approx 0.5$ mJ cm$^{-2}$ seeded by a Ti:Sapphire oscillator (Coherent Mira 900). The THz emission was detected via electro-optical sampling using a ZnTe nonlinear crystal. In the sample rotation measurements, a wire grid polarized was used to set the polarization plane of the detected THz radiation. All measurements were performed at room temperature.

**References**

[1] Sengupta, K., Nagatsuma, T. & Mittleman, D.M. Terahertz integrated electronic and hybrid electronic–photonic systems. *Nat. Electron.* **1,** 622–635 (2018).

[2] Fülöp, J.-A., Tzortzakis, S. & Kampfrath, T. Laser-driven strong-field terahertz sources. *Adv. Opt. Mater.* **8**, 1900681 (2019).

[3] Shumyatsky, P. & Alfano, R. R. Terahertz sources. *J. Biomed. Opt.* **16**, 033001 (2011).

[4] Seifert, T., Jaiswal, S., Martens, U. *et al.* Efficient metallic spintronic emitters of ultrabroadband terahertz radiation. *Nature Photon.* **10,** 483–488 (2016).

[5] Seifert, T. S., Jaiswal, S., Barker, J. *et al.* Femtosecond formation dynamics of the spin Seebeck effect revealed by terahertz spectroscopy. *Nat. Commun.* **9,** 2899 (2018).

[6] Wen, B., Ban, D High-temperature terahertz quantum cascade lasers. *Prog. Quantum Electron.* **80**, 100363 (2021).

[7] Lengfellner, H., Kremb, G., Schnellbögl, A., Betz, J., Renk, K. F., & Prettl, W. Giant voltages upon surface heating in normal YBa$_2$Cu$_3$O$_{7-\delta}$ films suggesting an atomic layer thermopile. *Appl. Phys. Lett.* **60**, 501–503 (1992).

[8] Lengfellner, H., Zeuner, S., Prettl, W., & Renk, K. F. Thermoelectric effect in normal-state YBa$_2$Cu$_3$O$_{7-\delta}$ films. *Europhys. Lett.* **25**, 375 (1994).

[9] Testardi, Louis R. Anomalous laser-induced voltages in YBa$_2$Cu$_3$O$_x$ and "off-diagonal" thermoelectricity. *Appl. Phys. Lett.* **64**, 2347 (1994).

[10] Zeuner, S., Lengfellner, H., & Prettl, W. Thermal boundary resistance and diffusivity for YBa$_2$Cu$_3$O$_{7-\delta}$ films. *Phys. Rev. B* **51**, 11903 (1995).

[11] Zahner, Th., Stierstorfer, R., Reindl, S., Schauer, T., Penzkofer, A., Lengfellner, H. Picosecond thermoelectric response in thin YBa$_2$Cu$_3$O$_{7-\delta}$ films. *Physica C* **313**, 37 (1999).

[12] Goldsmid, H. J., Application of the transvers thermoelectric effects. *J. Electron. Matter*. **40**, 1254 (2011).

[13] Zahner, Th., Schreiner, R., Stierstorfer, R., *et al.* Off-diagonal Seebeck effect and anisotropic thermopower in Bi$_2$Sr$_2$CaCu$_2$O$_8$ thin films. *Europhys. Lett.* **40**, 673–678 (1997).

[14] Zhang, P. X., Lee, W. K., Zhang, G. Y. Time dependence of laser-induced thermoelectric voltages in La$_{1-x}$Ca$_x$MnO$_3$ and YBa$_2$Cu$_3$O$_{7-\delta}$ thin films. *Appl. Phys. Lett.* **81**, 4026 (2002).

[15] Zhang, P. X., & Habermeier, H.-U. Atomic layer thermopile materials: physics and applications. *J. Nanomater.* **2008**, 329601 (2008).

[16] Habermeier, H.-U., Li, X. H., Zhang, P.X., Leibold, B. Anisotropy of thermoelectric properties in La$_{2/3}$Ca$_{1/3}$MnO$_3$ thin films studied by laser-induced transient voltages. *Solid State Commun.* **110**, 473 (1999).

[17] Xiong, F., Zhang, H., Jiang, Z. M., & Zhang, P. X. Transverse laser-induced thermoelectric voltages in tilted La$_{2-x}$Sr$_x$CuO$_4$ thin films. *J. Appl. Phys.* **104**, 053118 (2008).

[18] Yu, L., Wang, Y., Zhang, P., Habermeier, H.-U. Ultrafast transverse thermoelectric response in c-axis inclined epitaxial La$_{0.5}$Sr$_{0.5}$CoO$_3$ thin films. *Phys. Status Solidi RRL* **7**, 180 (2013).

[19] Takahashi, K., Kanno, T., Sakai, A., Adachi, H., & Yamada, Y. Gigantic transverse voltage induced via off-diagonal thermoelectric effect in Ca$_x$CoO$_2$ thin films. *Appl. Phys. Lett.* **97**, 021906 (2010).

[20] Takahashi, K., Kanno, T., Sakai, A., Adachi, H., & Yamada, Y. Influence of interband transition on the laser-induced voltage in thermoelectric Ca$_x$CoO$_2$ thin films. *Phys. Rev. B* **83**, 115107 (2011).

[21] Shan, J. & Heinz, T. F. in *Ultrafast Dynamical Processes in Semiconductors* (ed. Tsen, K.-T.) 1–56 (Springer, 2004).

[22] Shannon, R. D., Rogers, D. B., & Prewitt, C. T. Chemistry of noble metal oxides. I. Synthesis and properties of ABO2 delafossite compounds. *Inorg. Chem.* **10**, 713 (1971).

[23] Takatsu, H., Yonezawa, S., Mouri, S., Nakatsuji, S., Tanaka, K., & Maeno, Y. Roles of high-frequency optical phonons in the physical properties of the conductive delafossite $PdCoO_2$. *J. Phys. Soc. Jpn.* **76**, 104701 (2007).

[24] Hicks, C. W., Gibbs, A. S., Mackenzie, A. P., Takatsu, H., Maeno, Y., & Yelland, E.A. Quantum oscillations and high carrier mobility in delafossite $PdCoO_2$. *Phys. Rev. Lett.* **109**, 116401 (2012).

[25] Moll, P. J. W. W., Kushwaha, P., Nandi, N., Schmidt, B. & Mackenzie, A. P. Evidence for hydrodynamic electron flow in $PdCoO_2$. *Science* **351**, 1061–1064 (2016).

[26] Daou, R., Frésard, R., Hébert, S., & Maignan, A. Large anisotropic thermal conductivity of the intrinsically two-dimensional metallic oxide $PdCoO_2$. *Phys. Rev. B* **91**, 041113 (2015).

[27] Ong, K. P., Singh, D. J., & Wu, P. Unusual transport and strongly anisotropic thermopower in $PtCoO_2$ and $PdCoO_2$. *Phys. Rev. Lett.* **104**, 176601 (2010).

[28] Gruner, M. E., Eckern, U., & Pentcheva, R. Impact of strain-induced electronic topological transition on the thermoelectric properties of $PtCoO_2$ and $PdCoO_2$. *Phys. Rev. B* **92**, 235140 (2015).

[29] Yordanov, P., Sigle, W., Kaya, P., Gruner, M. E., Pentcheva, R., Keimer, B., & Habermeier, H.-U. Large thermopower anisotropy in $PdCoO_2$ thin films. *Phys. Rev. Mater.* **3**, 085403 (2019).

[30] Yordanov, P., Gibbs, A. S., Kaya, P., Bette, S., Xie, W., Xiao, X., Weidenkaff, A., Takagi, H., & Keimer, B. High-temperature electrical and thermal transport properties of polycrystalline $PdCoO_2$. *Phys. Rev. Mater.* **5**, 015404 (2021).

[31] Geisler, B., Yordanov, P. *et al*. Tuning the thermoelectric properties of transition-metal oxide thin films and superlattices on the quantum scale *Phys. Status Solidi B*, 2100270 (2021).

[32] Harada, T. Thin-film growth and application prospects of metallic delafossites *Mater. Today Adv.* **11**, 100146 (2021).

[33] Homes, C. C., Khim, S., & Mackenzie, A. P. Perfect separation of intraband and interband excitations in $PdCoO_2$. *Phys. Rev. B* **99**, 195127 (2019).

[34] Löffler, T., Hahn, T., Thomson, M., Jacob, F., & Roskos, H. G., Large-area electro-optic ZnTe terahertz emitters. *Opt. Express* **13**, 5353 (2005).

[35] Blanchard, F., Razzari, L., *et al*. Generation of 1.5 µJ single-cycle terahertz pulses by optical rectification from a large aperture ZnTe crystal. *Opt. Express* **15**, 13212 (2007).

[36] Takahashi, K., Kanno, T., Sakai, A., Tamaki, H., Kusada, H., Yamada, Y. Terahertz radiation via ultrafast manipulation of thermoelectric conversion in thermoelectric thin films. *Adv. Opt. Mater.* **2**, 428-434 (2014).

[37] Malevich, V. L. *et al*. THz emission from semiconductor surfaces. *C. R. Phys.* **9**, 130 (2008).

[38] Nakamura, Y., & Uchida, S. Anisotropic transport properties of single-crystal $La_{2-x}Sr_xCuO_4$: Evidence for the dimensional crossover. *Phys. Rev. B* **47**, 8369 (1993).

[39] Yan, J-Q., Zhou, J-S., & Goodenough, J. B. Thermal conductivity of $La_{2-x}Sr_xCuO_4$ ($0.05 \leq x \leq 0.22$). *New J. Phys*. **6**, 143 (2004).

[40] Mousatov, C. H., & Hartnoll, S. A. Phonons, electrons and thermal transport in Planckian high $T_c$ materials. *npj Quantum Mater*. **6**, 81 (2021).

[41] Najafi, E., Ivanov, V., Zewail, A., & Bernardi, M. Super-diffusion of excited carriers in semiconductors. *Nat. Commun.* **8**, 15177 (2017).

[42] Gedik, N., Orenstein, J., Liang, R., Bonn, D. A., Hardy, W. N., Diffusion of nonequilibrium quasi-particles in a cuprate superconductor. *Science* **300**, 1410 (2003).

[43] Block, A., Principi, A., Hesp, N.C.H. *et al.* Observation of giant and tunable thermal diffusivity of a Dirac fluid at room temperature. *Nat. Nanotechnol.* (2021).

[44] Mueller, B. Y., & Rethfeld, B., Relaxation dynamics in laser-excited metals under nonequilibrium conditions. *Phys. Rev. B* **87**, 035139 (2013).

[45] Hwang, H., Iwasa, Y., Kawasaki, M. *et al.* Emergent phenomena at oxide interfaces. *Nature Mater.* **11,** 103–113 (2012).

**Acknowledgements**

We acknowledge financial support by Deutsche Forschungsgemeinschaft (DFG, German Research Foundation) - Project No. 107745057 - CRC/TRR 80, subproject G8.

**Data availability**

The data that support the findings of this work are available from the corresponding authors upon reasonable request.

**Competing interests**

Patent pending.

**Author contributions**

P.Y. and S.K. conceived the project. P.Y., G.C., and G.L. synthesized and characterized the thin films. P.Y. performed the LITV experiments and analyzed the data. T.P. and M.-J. K. built the setup for the optical experiments. T.P., M.-J. K. and S.K. performed the optical experiments and analyzed the data. P.Y. and S.K. devised the model and carried out the calculations. P.Y., T.P., B.K., and S.K. wrote the paper, with contributions from all authors. B.K. and S.K. supervised the project.

# Intense terahertz radiation via the transverse thermoelectric effect

*Petar Yordanov, Tim Priessnitz, Min-Jae Kim, Georg Cristiani, Gennady Logvenov,*

*Bernhard Keimer, and Stefan Kaiser*

**Transverse thermoelectric effect (TTE) and laser-induced thermoelectric voltage (LITV)**

The TTE and LITV occur in natural or artificial materials with anisotropic Seebeck coefficient of planar ($S_a = S_b \equiv S_{ab} \neq S_c$) or lower symmetry, positioned such that their main crystallographic axes $\vec{a}\vec{b}, \vec{c}$ are inclined at an angle $\theta \neq 0{,}90°$ with respect to the temperature gradient $\vec{\nabla}T$ generated by a laser beam. For normal incidence of the beam, $\vec{\nabla}T$ is antiparallel to the surface normal $\vec{n}$ (Fig. S1a,b) [1-10]. The TTE emerges from off-diagonal elements in the Seebeck tensor, $\vec{E} = \boldsymbol{S} \cdot \vec{\nabla}T$, where $\vec{E}$ is the thermoelectric field.

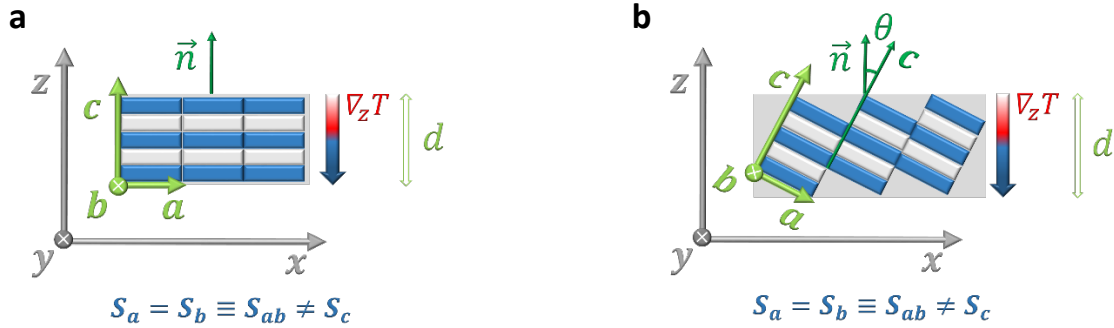

$$\boldsymbol{S} = \begin{pmatrix} S_{ab} & 0 & 0 \\ 0 & S_{ab} & 0 \\ 0 & 0 & S_c \end{pmatrix} \qquad \boldsymbol{S} = \begin{pmatrix} S_{ab}\cos^2\theta + S_c\sin^2\theta & 0 & (S_{ab}-S_c)\dfrac{\sin 2\theta}{2} \\ 0 & S_{ab} & 0 \\ (S_{ab}-S_c)\dfrac{\sin 2\theta}{2} & 0 & S_{ab}\sin^2\theta + S_c\cos^2\theta \end{pmatrix}$$

**Fig. S1: Transverse thermoelectric effect. a,** Layered material with anisotropic Seebeck coefficient positioned with the $\vec{a}\vec{b}, \vec{c}$ axes at angle $\theta = 90°, 0°$ relative to $\vec{n}$ and the temperature gradient $\nabla_z T$, respectively. The Seebeck tensor has only diagonal elements. **b,** Material positioned with the $\vec{a}\vec{b}, \vec{c}$ axes at angles $\theta \neq 90°, 0°$ relative to $\vec{n}$ and $\nabla_z T$. The Seebeck tensor has non-zero off-diagonal elements.

The thermoelectric field components are:

$$\begin{pmatrix} E_x \\ E_y \\ E_z \end{pmatrix} = \begin{pmatrix} S_{ab}\cos^2\theta + S_c\sin^2\theta & 0 & (S_{ab}-S_c)\frac{\sin 2\theta}{2} \\ 0 & S_{ab} & 0 \\ (S_{ab}-S_c)\frac{\sin 2\theta}{2} & 0 & S_{ab}\sin^2\theta + S_c\cos^2\theta \end{pmatrix} \cdot \begin{pmatrix} 0 \\ 0 \\ \nabla_z T \end{pmatrix}$$

$$= \begin{pmatrix} \frac{\sin 2\theta}{2}(S_{ab}-S_c)\nabla_z T \\ 0 \\ (S_{ab}\sin^2\theta + S_c\cos^2\theta)\nabla_z T \end{pmatrix} \quad (S1)$$

where $E_x$ describes the TTE, $E_y$ vanishes by symmetry, and $E_z$ is due to the ordinary longitudinal Seebeck effect.

In LITV experiments on films with thickness much smaller than the laser spot diameter $d \ll l$, the heat flow is practically always perpendicular to the film surface (Fig. S2). The corresponding voltage components are:

$$U_x = E_x l = \frac{l}{d}\frac{\sin 2\theta}{2}\Delta S \Delta T_z \quad (S2)$$

$$U_y = E_y l = 0 \quad (S3)$$

$$U_z = E_z d = (S_{ab}\sin^2\theta + S_c\cos^2\theta)\Delta T_z \quad (S4)$$

where $\Delta S = (S_{ab} - S_c)$, $\nabla_z T = \Delta T_z/d$, and $\Delta T_z = T_f - T_b$ is the temperature difference between the front and the back sides of the film.

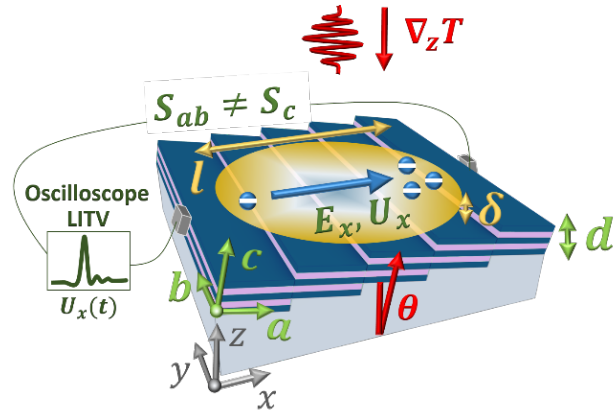

**Fig. S2: Laser-induced thermoelectric voltage.** Thin film of a material with anisotropic Seebeck coefficient deposited on an offcut substrate, which forces the material axes $\vec{ab}, \vec{c}$ to grow at angles $\theta \neq 90,0°$ relative to the surface normal and the temperature gradient. In LITV, the $U_x(t) \sim \Delta T_z(t)$ dependence is studied and the signal $U_x(t)$ is detected by means of electrical measurements (oscilloscope).

The defining feature of the TTE/LITV effects, $U_x \perp \Delta T_z$, and the aspect ratio $l/d$ ($d \approx 10 - 100$ nm, $l \approx 1 - 10$ mm), which introduces a signal enhancement of several orders of magnitude, enable highly sensitive experiments and studies of the thermoelectric conversion dynamics in films through the relation $U_x(t) \sim \Delta T_z(t)$, because the time necessary for heat to travel across the film is very short. This may include investigations of the time-dependent voltage $U_x(t)$, as well as the TTE associated with the electric current. The TTE thus allows access to one of the less-studied aspects of the thermoelectric conversion mechanism, namely its dynamical behavior and related functionalities. On the microscopic scale, the TTE-voltage buildup can be viewed as activation of a colossal number of atomic-layer thermocouples in series formed by the material's layered structure [1,8]. Larger inclination angles $\theta$ (up to 45°) provide a better exposure to the heat flux. The development of the TTE voltage is an extremely fast process, because the distances between these "discrete thermocouples" are of the order of the unit cell, and hence, the times necessary for the carriers to travel are very short. However, the voltage $U_x(t)$ will eventually follow $\Delta_z T(t)$, which is described by the one-dimensional heat equation $\partial T(z,t)/\partial t = D\,\partial^2 T(z,t)/\partial z^2$, where $D = \kappa/\rho_m c_p$ is the thermal diffusivity, $\rho_m$ is the density, and $c_p$ the specific heat of the material. Thermal models based on the heat equation and Eq. S2, have established the basis for the description of the experimentally observed LITV signals [4,9,10]. Essentially, the characteristic response time is determined by the time necessary for thermal diffusion across the film thickness $\tau_r \approx 0.4\,d^2/D_z$ [4]. The response time $\tau_r$ can thus be reduced by decreasing the film thickness and/or by selecting materials with a large thermal diffusivity. The thickness, however, cannot be arbitrarily small since this may deteriorate the temperature drop $\Delta T_z$ across the sample. Similar arguments apply to the thermal diffusivity, that is, materials with larger thermal diffusivity will exhibit shorter response times, but at the expense of some signal strength. The optimization for large peak voltage requires strong thermopower anisotropies, as well as maximum laser fluence $F_{lp}$ - to enhance $\Delta T_z$, preferably at frequencies corresponding to the material optical absorption bands. In this case, the thermal stability becomes an important factor. Many layered transition metal oxides (TMO) comprising alternating metallic and insulating layers, exhibit thermopower anisotropies and are thermally stable. LITV experiments with nano- and sub-nanosecond laser pulses on thin films of TMOs, such as $YBa_2Cu_3O_{7-x}$ [1-5,9], $Bi_2Sr_2CaCu_2O_8$ [7], $La_{1-x}Ca_xMnO_3$ [9,11], $La_{2-x}Sr_xCuO_4$ ($x = 0.05 - 0.3$) [12], $La_{0.5}Sr_{0.5}CoO_3$ [13], $Ca_xCoO_2$ [10,14], and others, revealed the generation of large-amplitude (of up to several tens of volts) and fast (from micro- to sub-nanosecond) $U_x(t)$ signals.

**Time-dependent model for $U_x$**

An expression for $U_x(t)$ that assumes instantaneous heating of a layer of thickness $\delta_{opt,z}$ (a layer of uniform temperature determined by the optical penetration depth) (Fig. S2) at time $t \to 0$ was introduced in Ref. [9]:

$$U_x(t) = \frac{l}{d}\frac{\sin 2\theta}{2}(S_{ab} - S_c) \times \frac{F_{lp}A_{opt,z}}{\rho_m c_p \sqrt{\pi 4 D_z t}}\left(e^{-\frac{\delta_{opt,z}^2}{4D_z t}} - e^{-\frac{d^2}{4D_z t}}\right) \quad (S5)$$

In Eq. S5, the first term represents the "intrinsic" film effective Seebeck coefficient, while the second term yields the time-dependent temperature difference $\Delta T_z(t)$. $F_{lp}$ is the laser fluence, and $A_{opt,z}$ the optical absorption of the film.

**Experimental and model $U_x(t)$ of PdCoO$_2$ thin films**

Since $U_x$ is proportional to the difference $\Delta S = S_{ab} - S_c$, the LITV can be used to confirm and quantify the thermopower anisotropy. Fig. S3a shows typical experimental $U_x(t)$ data and model calculations for a $d \approx 10$ nm PdCoO$_2$ film deposited on a $\theta = 10°$ offcut Al$_2$O$_3$ (0001) → (11$\bar{2}$0) substrate, obtained by illumination with an excimer laser (wavelength $\lambda_{lp} = 248$ nm, pulse duration $\tau_{lp} \approx 5$ ns, repetition rate 1 Hz, and fluence $F_{lp} \approx 70$ mJ cm$^{-2}$). The data demonstrates a strong $U_x(t)$ signal, with a peak voltage $U_x \approx 18$ V, and interestingly, a full width at half maximum (FWHM) of ~7 ns, which is comparable to $\tau_{lp}$. The result confirms the large thermopower anisotropy in PdCoO$_2$ [15-19], whereas the absence of noticeable broadening indicates that the intrinsic response time $\tau_r$ of the PdCoO$_2$ films is likely shorter than $\tau_{lp}$.

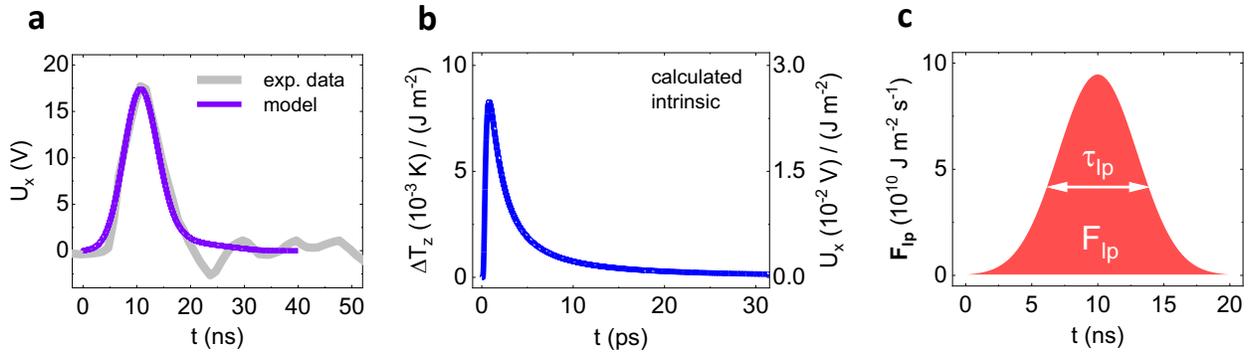

**Fig. S3: Laser-induced thermoelectric voltage of PdCoO$_2$ thin films. a** Experimental and model $U_x(t)$ of $d \approx 10$ nm PdCoO$_2$ thin film deposited on $\theta = 10°$ offcut substrate, obtained by an excimer laser ($\lambda_{lp} = 248$ nm), with $\tau_{lp} \approx 5$ ns, $RR_{lp} = 1$ Hz, and $F_{lp} \approx 70$ mJ cm$^{-2}$. **b** Calculated film intrinsic $\Delta T_z(t)$ and $U_x(t)$ (per fluence $F_{lp} = 1$ J m$^{-2}$) for $d = 10$ nm, $\theta = 10°$ PdCoO$_2$ film. **c** Gaussian profile $\boldsymbol{F_{lp}}(t)$ representing the laser pulse characteristics, with a FWHM of 6 ns = $\tau_{lp}$, and area 71.3 mJ cm$^{-2}$ = $F_{lp}$.

We use Eq. S5 to calculate the intrinsic $U_x(t)$ of the PdCoO$_2$ films and to fit the experimental data. The model parameters (at 300 K) read; $l = 5$ mm, $\theta = 10°$, $d = 10$ nm, $\Delta S = 34$ μV K$^{-1}$, $\rho_m = 7.99$ g cm$^{-3}$, $c_p = 350$ J kg$^{-1}$ K$^{-1}$, $D_{ab} = 1.1 \times 10^{-4}$ m$^2$ s$^{-1}$ and $D_c = 1.8 \times 10^{-5}$ m$^2$ s$^{-1}$ [17-20]. For $A_{opt,z}$, we assume the optical absorptivity $A_{opt,z} = 1 - R_{opt,z} - T_{opt,z}$, where $R_{opt,z}$ and $T_{opt,z}$ are the reflectivity and transmittivity of the film, with absorption coefficient $\alpha_{opt,z}$ and thickness $d$. At $\lambda_{lp} = 248$ nm, and a film thickness of $d = 10$ nm, we estimate roughly $A_{opt,ab} \approx A_{opt,c} \approx 0.4$, and $\delta_{opt,ab} \approx \delta_{opt,c} \approx 40$ nm, based on a model for the dielectric function and experimental transmission spectra [21,22]. The anisotropic quantities are introduced with their effective values along the $z$ axis; $A_{opt,z} = A_{opt,ab}\cos^2\theta + A_{opt,c}\sin^2\theta$, $\delta_{opt,z} = \delta_{opt,ab}\cos^2\theta + \delta_{opt,c}\sin^2\theta$, and $D_z = D_{ab}\sin^2\theta + D_c\cos^2\theta$. Since for films of thickness $d = 10$ nm, $\delta_{opt,z} > d$ is always fulfilled, we use the penetration depth to fit the experimental voltage (essentially a fit of the difference $\Delta T_z$) by $\delta_{opt,z}^{fit} = d - f$, where $f$ is the fit parameter ($f \ll d$). Indeed, as the substrate material is optically transparent and does not increase its temperature upon laser illumination, the use of the full penetration depth $\delta_{opt,z}$ is inappropriate in the case of $\delta_{opt,z} > d$. The parameter $f$ could, in principle, be related to the existence of a film-substrate interface layer, with properties different from those of the film. Alternatively, other mechanisms such as heat dissipation in the lateral direction can be effective in maintaining the difference $\Delta T_z$. The model result (Fig. S3a) is obtained by the convolution of the film-intrinsic $U_x(t)$ (normalized to a fluence of $1$ J m$^{-2}$, Fig. S3b) with a Gaussian $F_{lp}(t)$ (Fig. S3c) representing the laser pulse characteristics [9], namely $FWHM = \tau_{lp} = 5 - 7$ ns (taking into account broadening), and Area $= F_{lp} = 71.3$ mJ cm$^{-2}$ (Eq. S6):

$$U_x(t) = \frac{l}{d}\frac{sin2\theta}{2}(S_{ab} - S_c) \times \frac{A_{opt,z}}{\rho_m c_p \sqrt{\pi 4 D_z t}} \left( e^{-\frac{\delta_{opt,z}^2}{4D_z t}} - e^{-\frac{d^2}{4D_z t}} \right) * F_{lp}(t) \times \tau_{lp} c_f \qquad (S6)$$

where $F_{lp}(t) = \frac{2F_{lp}}{\tau_{lp}\sqrt{\pi}} e^{-\frac{4t^2}{\tau_{lp}^2}}$, and $f(t') * F_{lp}(t) = \int_{-\infty}^{\infty} f(t') F_{lp}(t-t') \, dt$.

The fit of the experimental data is obtained for $f = 0.0125$ nm corresponding to a maximum temperature difference of $\Delta T_z = 5.82$ K. The final result, after the convolution, is multiplied by $\tau_{lp} c_f$, where $\tau_{lp}$ solves the units, and $c_f$ is a correction factor related to the convolution details such as sampling interval, data column lengths, etc.; $c_f = \sqrt{\pi}/(2 \times 4225)$. $c_f$ is determined by a consistency check of $\Delta T_z(K)/(1$ J m$^{-2}) \times F_{lp}$ with the final result for the maximum of $\Delta T_z(K)$ after the convolution with $F_{lp}(t)$. We obtain an excellent qualitative model description of the experimental data (Fig. S3a). Clearly, in the case of $\delta_{opt,z} > d$, a quantitative model description would require further work on the definition of the fit parameter $f$ and the correction factor $c_f$. Nonetheless, the obtained experimental and model results directly confirm the large thermopower anisotropy in PdCoO$_2$. The calculated intrinsic $U_x(t)$ (Fig. S3b) reveals a remarkably short intrinsic response time $\tau_r$ of the $d = 10$ nm PdCoO$_2$ films, with a FWHM of 2.2 ps.

## Generation of terahertz radiation via the transverse thermoelectric effect

The buildup of thermoelectric field $E_x(t)$ and voltage $U_x(t)$ upon pulsed laser illumination is realized through thermally driven charge carrier diffusion, equivalent to an electric current $I_x(t)$ and current density $j_x(t)$ (Fig. S4). Any change of $j_x(t)$ with time (accelerated / decelerated charge carriers) can thus give rise to terahertz (THz) dipole radiation, $E_{\text{THz},z} \sim \partial j_x(t)/\partial t$, if the material's response time is in the range $\tau_r \sim d^2/D_z \lesssim 10^{-12}$ s and sufficiently short laser pulses are used $\tau_{\text{lp}} \lesssim \tau_r$.

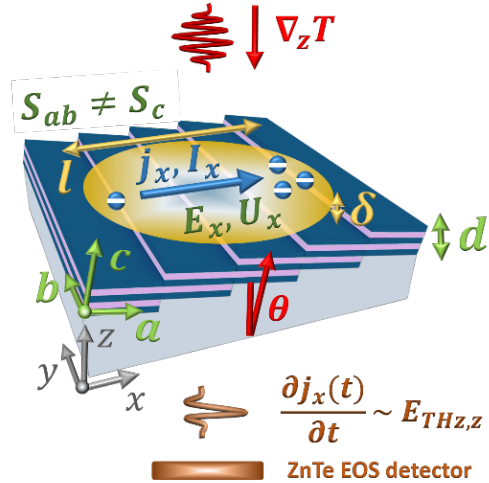

**Fig. S4: TTE-THz generation.** Thin film of material with anisotropic Seebeck coefficient on offcut substrate $\theta \neq 0, 90°$ is illuminated with a femtosecond laser pulse. The THz radiation results from a time-varying current density $E_{\text{THz},z} \sim \partial j_x(t)/\partial t$. In this work, $E_{\text{THz},z}$ is detected by electro-optical sampling (EOS) with a ZnTe crystal.

By utilizing terahertz emission spectroscopy, in which we make use of femtosecond ($\tau_{\text{lp}} = 250$ fs, $\lambda_{\text{lp}} = 800$ nm) laser pulses as sources of ultrafast temperature gradients, we generated a linearly polarized THz emission in the range $\nu \approx 0.1 - 2.3$ THz from thin films of PdCoO$_2$ and La$_{2-x}$Sr$_x$CuO$_4$ ($x = 0.16$) (LSCO) deposited on offcut substrates, at room temperature.

## Model description of the TTE-THz generation

Since we assume that time-varying currents associated with the thermoelectric voltage dynamics are the sources of the THz radiation, a combination of equations for the voltage $U_x(t)$ with an equation for the current $\partial j_x(t)/\partial t$ should be able to provide a consistent description of the TTE-THz field $E_{\text{THz},z}$. We combine Eq. S5 with the expression for the far-field zone used in the description of photoconductive THz emitters (Eq. S7) [23]:

$$E_{\text{THz}}(t) \cong -\frac{2A}{c^2 r}\frac{1}{1+\sqrt{\varepsilon}}\frac{\partial j(t)}{\partial t} \qquad (S7)$$

where $A$ is the illuminated surface area in which the current density is present, $r$ is the distance of propagation, $\varepsilon = n^2$ refer to the real dielectric function and refractive index at THz frequencies, respectively (dispersion is neglected), $c$ is the speed of light, and $j(t)$ is the current density.

In Eq. S7, we insert the current $j_x(t) = I_x(t)/ld \approx U_x(t)/R_x ld$, where $U_x(t)$ is the result from Eq. S5, and $I_x(t)$ is calculated based on the $dc$-resistance $R_x$; $I_x(t) \approx U_x(t)/R_x$. Further, Eq. S7 is adapted for currents in a volume $V = \pi(l/2)^2 d$, and multiplied by $1/4\pi\varepsilon_0$ for unit conversion, where $\varepsilon_0$ is the dielectric permittivity of free space. Similarly to Eq. S6, the laser pulse is introduced at the last step, by convolution with a Gaussian profile with area $F_{\text{lp}}$ and FWHM $\tau_{\text{lp}}$:

$$E_{\text{THz},z}(t) \approx -\frac{2V}{4\pi\varepsilon_0 c^2 r_z}\frac{1}{(1+n_{\text{opt},z})}\frac{\partial j_x(t)}{\partial t} * \boldsymbol{F_{\text{lp}}}(t) \times \tau_{\text{lp}} c_f \qquad (S8)$$

The anisotropic quantities are introduced with their effective values along the respective direction; $R_x = R_{ab}\cos^2\theta + R_c\sin^2\theta$, $n_{\text{opt},z} = n_{\text{opt},ab}\cos^2\theta + n_{\text{opt},c}\sin^2\theta$, and $r_z = d$.

A modification of the thermal diffusivity $D_z$ in Eq. S5 that takes into account the transient state $D_z^*(t)$ (resulting from the illumination with the laser pulse), is crucial for the model description of $E_{\text{THz},z}(t)$. $D_z^*(t)$ may reach values several orders of magnitude larger compared to equilibrium $D_z^* \gtrsim 1000 \times D_z^0$, and is particularly important to consider in cases where the pump laser pulse duration $\tau_{lp}$ (the temporal resolution) is shorter than the time for complete recovery $D_z^* \to D_z^0$. In metals, the characteristic thermalization times $\tau_{\text{th}}$ are short – of the order of a few tens to a few hundreds of femtoseconds, depending on the laser pulse details (typically $\tau_{\text{th}} \approx 100$ fs), and the system relaxes within a few picoseconds [24-27]. In our model, we define a time-dependent thermal diffusivity of the form:

$$D_z(t) = D_z^0 + (D_z^* - D_z^0)e^{-\frac{t}{\tau_{\text{th}}}} \qquad (S9)$$

where we allow a transient diffusivity of up to $D_z^* \approx 5000 \times D_z^0$, and a thermalization time in the range $\tau_{\text{th}} \approx 70 - 120$ fs (Fig. S5a). As a consequence, the temporal profiles of $D_z(t)$, $\Delta T_z(t)$, $j_x(t)$, $E_{\text{THz},z}(t)$, and the corresponding spectra, consist of a faster transient and a slower equilibrium component (Fig. S5a-f). We note that Eq. S9 yields an excellent qualitative description of the $E_{\text{THz},z}(t)$ waveforms, and explains the pronounced gap in the THz spectrum of the LSCO films at ~0.5 THz. It also suggests a potential bandwidth tuning parameter in the TTE-THz generation, $D_z(F_{\text{lp}}, \tau_{\text{lp}})$, which is worth further investigations.

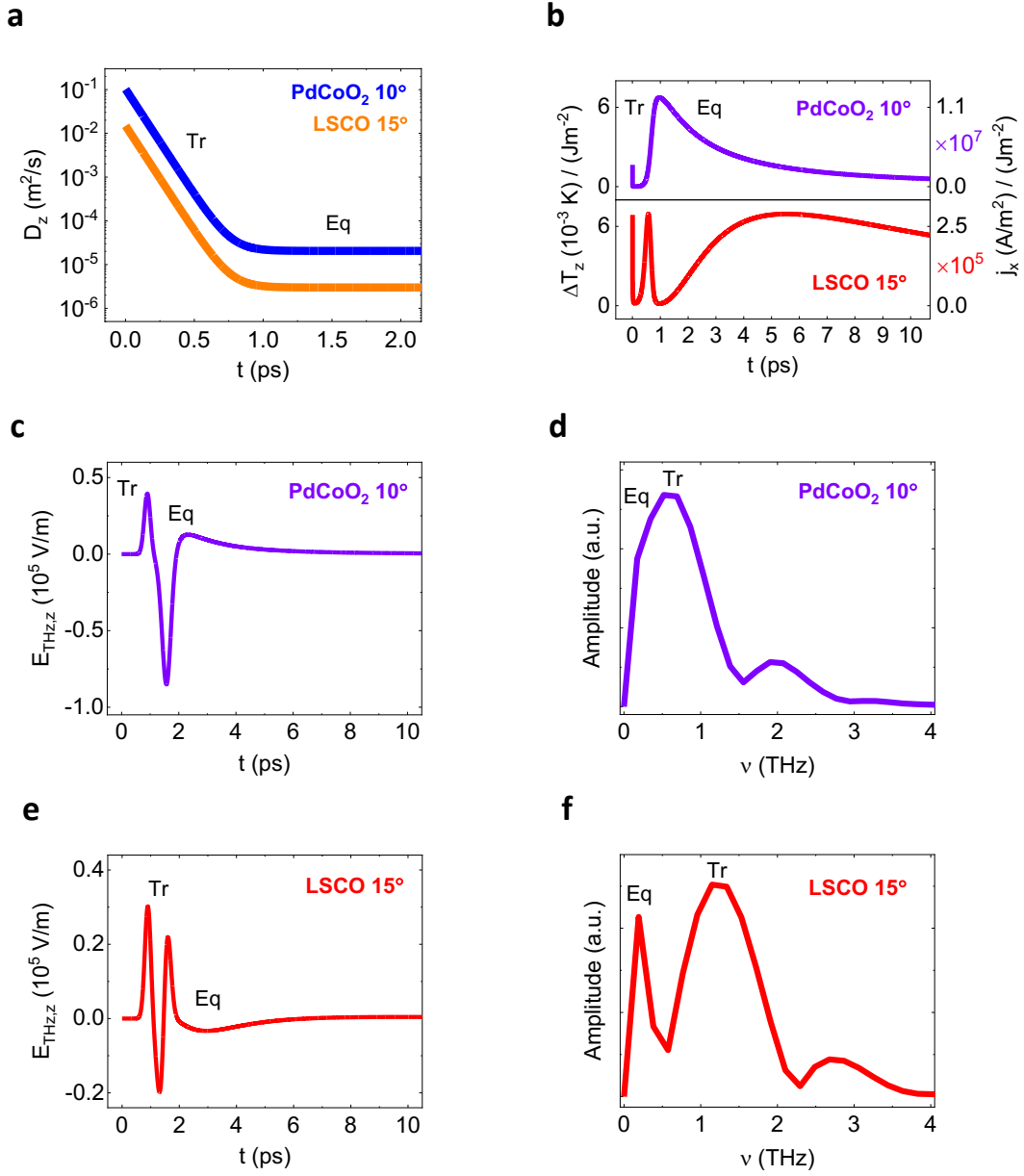

**Fig. S5: TTE-THz model description.** Model results for $d = 10$ nm $\theta = 10°$ PdCoO$_2$ and $d = 10$ nm $\theta = 15°$ LSCO films; **a,** Time-dependent thermal diffusivity $D_z(t)$ (log scale). **b** Film-intrinsic temperature difference $\Delta T_z(t)$ and current density $j_x(t)$ profiles (per $1 \, \text{J m}^{-2}$ fluence). **c,d,** and **e,f,** $E_{\text{THz},z}(t)$ waveforms and amplitude spectra. Due to the compound thermal diffusivity $D_z(t)$, $\Delta T_z(t)$, $j_x(t)$, $E_{\text{THz},z}(t)$ and the corresponding spectra, consist of transient and equilibrium components, which are denoted by (Tr) and (Eq), respectively.

For completeness, we list the input model parameters for PdCoO$_2$ films (Fig. S5 c,d): $l = 1$ mm, $\theta = 10°$, $d = 10$ nm, $\Delta S = 34$ μV K$^{-1}$, $\rho_m = 7.99$ g cm$^{-3}$, $c_p = 350$ J kg$^{-1}$ K$^{-1}$, $A_{\text{opt},ab} \approx 0.35$, $A_{\text{opt},c} \approx 0.13$, $\delta_{\text{opt},ab} \approx 70$ nm, $\delta_{\text{opt},c} \approx 100$ nm (at $\lambda_{\text{lp}} = 800$ nm), $f = 1.25 \times 10^{-11}$ (fixed from the fit of the LITV data), $D_{ab} = 1.1 \times 10^{-4}$ m$^2$ s$^{-1}$, $D_c = 1.8 \times 10^{-5}$ m$^2$ s$^{-1}$, $D_z(t) = D_z^0 + (D_z^* - D_z^0)e^{-\frac{t}{\tau_{\text{th}}}}$, $D_z^* = 5000 \times D_z^0$, $\tau_{\text{th}} = 90$ fs, $n_{\text{opt},ab} = 330$, $n_{\text{opt},c} = 37$ (at frequency $\nu = 1$ THz). The optical quantities were estimated from a dielectric function model, with plasma frequency $\omega_{\text{p},ab} = 33317$ cm$^{-1}$ and $\omega_{\text{p},c} = 3660$ cm$^{-1}$, and from experimental transmission spectra. $R_{ab} = 3$ μΩ cm/d, $R_c = 1000$ μΩ cm/d [17-22, 28,29]. The laser pulse is represented by a Gaussian profile $\boldsymbol{F}_{\text{lp}}(t)$ with $FWHM = \tau_{\text{lp}} = 250$ fs, and $Area = F_{\text{lp}} = 5$ J m$^{-2}$, $\tau_{\text{lp}} = 250$ fs. $c_f = \sqrt{\pi}/(2 \times 237)$, where $c_f$ is determined by comparison of the peak $E_{\text{THz},z}$ with that of ZnTe optical rectification source [30,31], and by consistency check of $\Delta T_z(K)/(1 \text{ J m}^{-2}) \times F_{\text{lp}}$ with the final result for the maximum of $\Delta T_z(K)$ after the convolution with $\boldsymbol{F}_{\text{lp}}(t)$. The resulting maximum temperature difference is $\Delta T_z = 37$ mK, and the current density $j_x = 1.22 \times 10^7$ A m$^{-2}$, at $F_{\text{lp}} = 0.5$ mJ cm$^{-2}$.

For LSCO ($x = 0.16$) films (Fig. S5 e,f), the following input parameters were used: $\theta = 15°$, $\Delta S = 40$ μV K$^{-1}$, $D_{ab} \approx D_c = 3 \times 10^{-6}$ m$^2$ s$^{-1}$, $R_{ab} = 150$ μΩ cm/d, $R_c = 30000$ μΩ cm/d, and $n_{\text{opt},ab} = 33$, $n_{\text{opt},c} = 3.7$ (at $\nu = 1$ THz) [12, 32-36]. The remaining parameters are identical to those for PdCoO$_2$.

### $E_{\text{THz},z}(\theta)$ and $E_{\text{THz},z}(d)$ dependence and optimal film parameters

The anisotropic nature of the material parameters and the impact of the film thickness on the temperature drop $\Delta T_z$ imply the existence of an optimal offcut angle and thickness, $E_{\text{THz},z}(\theta)$ and $E_{\text{THz},z}(d)$. The relevant parameters include the in-plane resistivity $\rho_x$, which determines the magnitude of $j_x$; the optical absorptivity $A_{\text{opt},z}$ and penetration depth $\delta_z$ which are decisive for the magnitude of $\Delta T_z$; and the refractive index $n_z$, which is responsible for the out-coupling of THz radiation. In addition, the out-of-plane thermal diffusivity $D_z$ influences both the operating speed and the signal amplitude, that is, higher $D_z$ leads to shorter $\tau_r$ (larger bandwidth) at the expense of $\Delta T_z$. As already discussed, the film thickness $d$ has a similar but reciprocal effect. The calculations yield maxima of $E_{\text{THz},z}$ at small offcut angles $\theta \approx 3 - 7°$, and for film thickness between $d \approx 110 - 170$ nm (in regime $d > \delta_{\text{opt},z}$) where $\delta_{\text{opt},ab} = 70$ nm, and $\delta_{\text{opt},c} = 100$ nm (Fig S6a,b), given that the transient thermal diffusivity $D_z^* > 1000 \times D_z^0$ is indeed realized.

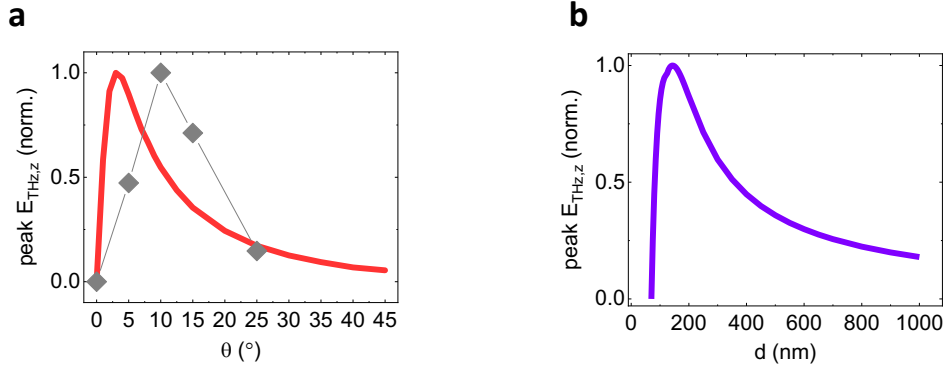

**Fig. S6: $E_{\text{THz},z}(\theta)$ and $E_{\text{THz},z}(d)$ dependence: model results for PdCoO$_2$ films (300 K). a,** Calculated peak field $E_{\text{THz},z}$ (solid line) for $d > \delta_{\text{opt},z}$ and experimental data (connected symbols) for $d \approx 10$ nm PdCoO$_2$ films (normalized data) as a function of the offcut angle. **b,** Calculated peak field $E_{\text{THz},z}$ as a function of the film thickness in regime $d > \delta_{\text{opt},z}$ (normalized data).

The comparison between the calculated and experimental $E_{\text{THz},z}(\theta)$ data in Fig. S6a shows overall good agreement. The discrepancy is likely due to differences between the anisotropy ratios of the model input parameters, which were taken from bulk measurements, and those of the films used in the experiments.

**Simulation with optimal film parameters**

A simulation with film parameters; $\theta = 3°$, $d = 110\ nm$ ($d > \delta_{\text{opt},z}$), $D_z^* = 5000 \times D_z^0$, and ultrashort laser pulse $\tau_{\text{lp}} = 10$ fs, indicate that PdCoO$_2$ has the potential to generate intense THz radiation with peak field of $E_{\text{THz},z} \approx 1 \times 10^6$ V m$^{-1}$ at $F_{\text{lp}} = 0.5$ mJ cm$^{-2}$, and bandwidth of $\sim 0.1 - 50$ THz (Fig. S7).

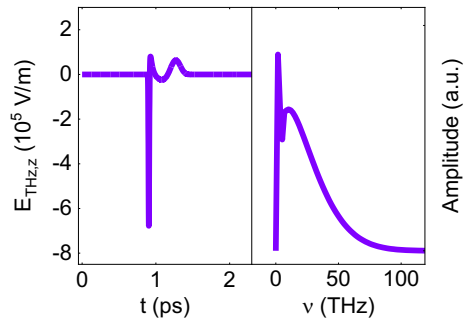

**Fig. S7:** Simulated $E_{\text{THz},z}(t)$ and amplitude spectrum for optimal PdCoO$_2$ film parameters, pulse width $\tau_{\text{lp}} = 10$ fs and fluence $F_{\text{lp}} = 0.5$ mJ cm$^{-2}$.

## References

[1] Lengfellner, H., Kremb, G., Schnellbögl, A., Betz, J., Renk, K. F., & Prettl, W. Giant voltages upon surface heating in normal YBa$_2$Cu$_3$O$_{7-\delta}$ films suggesting an atomic layer thermopile. *Appl. Phys. Lett.* **60**, 501–503 (1992).

[2] Lengfellner, H., Zeuner, S., Prettl, W., & Renk, K. F. Thermoelectric effect in normal-state YBa$_2$Cu$_3$O$_{7-\delta}$ films. *Europhys. Lett.* **25**, 375 (1994).

[3] Testardi, Louis R. Anomalous laser-induced voltages in YBa$_2$Cu$_3$O$_x$ and "off-diagonal" thermoelectricity. *Appl. Phys. Lett.* **64**, 2347 (1994).

[4] Zeuner, S., Lengfellner, H., & Prettl, W. Thermal boundary resistance and diffusivity for YBa$_2$Cu$_3$O$_{7-\delta}$ films. *Phys. Rev. B* **51**, 11903 (1995).

[5] Zahner, Th., Stierstorfer, R., Reindl, S., Schauer, T., Penzkofer, A., Lengfellner, H. Picosecond thermoelectric response in thin YBa$_2$Cu$_3$O$_{7-\delta}$ films. *Physica C* **313**, 37 (1999).

[6] Goldsmid, H. J., Application of the transvers thermoelectric effects. *J. Electron. Matter.* **40**, 1254 (2011).

[7] Zahner, Th., Schreiner, R., Stierstorfer, R., *et al.* Off-diagonal Seebeck effect and anisotropic thermopower in Bi$_2$Sr$_2$CaCu$_2$O$_8$ thin films. *Europhys. Lett.* **40**, 673–678 (1997).

[8] Zhang, P. X., & Habermeier, H.-U. Atomic layer thermopile materials: physics and applications. *J. Nanomater.* **2008**, 329601 (2008).

[9] Zhang, P. X., Lee, W. K., Zhang, G. Y. Time dependence of laser-induced thermoelectric voltages in La$_{1-x}$Ca$_x$MnO$_3$ and YBa$_2$Cu$_3$O$_{7-\delta}$ thin films. *Appl. Phys. Lett.* **81**, 4026 (2002).

[10] Takahashi, K., Kanno, T., Sakai, A., Adachi, H., & Yamada, Y. Influence of interband transition on the laser-induced voltage in thermoelectric Ca$_x$CoO$_2$ thin films. *Phys. Rev. B* **83**, 115107 (2011).

[11] Habermeier, H.-U., Li, X. H., Zhang, P.X., Leibold, B. Anisotropy of thermoelectric properties in La$_{2/3}$Ca$_{1/3}$MnO$_3$ thin films studied by laser-induced transient voltages. *Solid State Commun.* **110**, 473 (1999).

[12] Xiong, F., Zhang, H., Jiang, Z. M., & Zhang, P. X. Transverse laser-induced thermoelectric voltages in tilted La$_{2-x}$Sr$_x$CuO$_4$ thin films. *J. Appl. Phys.* **104**, 053118 (2008).

[13] Yu, L., Wang, Y., Zhang, P., Habermeier, H.-U. Ultrafast transverse thermoelectric response in c-axis inclined epitaxial La$_{0.5}$Sr$_{0.5}$CoO$_3$ thin films. *Phys. Status Solidi RRL* **7**, 180 (2013).

[14] Takahashi, K., Kanno, T., Sakai, A., Adachi, H., & Yamada, Y. Gigantic transverse voltage induced via off-diagonal thermoelectric effect in $Ca_xCoO_2$ thin films. *Appl. Phys. Lett.* **97**, 021906 (2010).

[15] Ong, K. P., Singh, D. J., & Wu, P. Unusual transport and strongly anisotropic thermopower in $PtCoO_2$ and $PdCoO_2$. *Phys. Rev. Lett.* **104**, 176601 (2010).

[16] Gruner, M. E., Eckern, U., & Pentcheva, R. Impact of strain-induced electronic topological transition on the thermoelectric properties of $PtCoO_2$ and $PdCoO_2$. *Phys. Rev. B* **92**, 235140 (2015).

[17] Yordanov, P., Sigle, W., Kaya, P., Gruner, M. E., Pentcheva, R., Keimer, B., & Habermeier, H.-U. Large thermopower anisotropy in $PdCoO_2$ thin films. *Phys. Rev. Mater.* **3**, 085403 (2019).

[18] Yordanov, P., Gibbs, A. S., Kaya, P., Bette, S., Xie, W., Xiao, X., Weidenkaff, A., Takagi, H., & Keimer, B. High-temperature electrical and thermal transport properties of polycrystalline $PdCoO_2$. *Phys. Rev. Mater.* **5**, 015404 (2021).

[19] Geisler, B., Yordanov, P. *et al*. Tuning the thermoelectric properties of transition-metal oxide thin films and superlattices on the quantum scale *Phys. Status Solidi B*, 2100270 (2021).

[20] Daou, R., Frésard, R., Hébert, S., & Maignan, A. Large anisotropic thermal conductivity of the intrinsically two-dimensional metallic oxide $PdCoO_2$. *Phys. Rev. B* **91**, 041113 (2015).

[21] Homes, C. C., Khim, S., & Mackenzie, A. P. Perfect separation of intraband and interband excitations in $PdCoO_2$. *Phys. Rev. B* **99**, 195127 (2019).

[22] Harada, T. *et al*. Highly conductive $PdCoO_2$ ultrathin films for transparent electrodes. *APL Mater*. **6**, 046107 (2018).

[23] Shan, J. & Heinz, T. F. in *Ultrafast Dynamical Processes in Semiconductors* (ed. Tsen, K.-T.) 1–56 (Springer, 2004).

[24] Najafi, E., Ivanov, V., Zewail, A., & Bernardi, M. Super-diffusion of excited carriers in semiconductors. *Nat. Commun.* **8**, 15177 (2017).

[25] Gedik, N., Orenstein, J., Liang, R., Bonn, D. A., Hardy, W. N., Diffusion of nonequilibrium quasi-particles in a cuprate superconductor. *Science* **300**, 1410 (2003).

[26] Mueller, B. Y., & Rethfeld, B., Relaxation dynamics in laser-excited metals under nonequilibrium conditions. *Phys. Rev. B* **87**, 035139 (2013).

[27] Block, A., Principi, A., Hesp, N.C.H. *et al.* Observation of giant and tunable thermal diffusivity of a Dirac fluid at room temperature. *Nat. Nanotechnol.* (2021).

[28] Takatsu, H., Yonezawa, S., Mouri, S., Nakatsuji, S., Tanaka, K., & Maeno, Y. Roles of high-frequency optical phonons in the physical properties of the conductive delafossite $PdCoO_2$. *J. Phys. Soc. Jpn.* **76**, 104701 (2007).

[29] Hicks, C. W., Gibbs, A. S., Mackenzie, A. P., Takatsu, H., Maeno, Y., & Yelland, E.A. Quantum oscillations and high carrier mobility in delafossite $PdCoO_2$. *Phys. Rev. Lett.* **109**, 116401 (2012).

[30] Löffler, T., Hahn, T., Thomson, M., Jacob, F., & Roskos, H. G., Large-area electro-optic ZnTe terahertz emitters. *Opt. Express* **13**, 5353 (2005).

[31] Blanchard, F., Razzari, L., *et al*. Generation of 1.5 μJ single-cycle terahertz pulses by optical rectification from a large aperture ZnTe crystal. *Opt. Express* **15**, 13212 (2007).

[32] Nakamura, Y., & Uchida, S. Anisotropic transport properties of single-crystal $La_{2-x}Sr_xCuO_4$: Evidence for the dimensional crossover. *Phys. Rev. B* **47**, 8369 (1993).

[33] Yan, J-Q., Zhou, J-S., & Goodenough, J. B. Thermal conductivity of $La_{2-x}Sr_xCuO_4$ ($0.05 \leq x \leq 0.22$). *New J. Phys*. **6**, 143 (2004).

[34] Mousatov, C. H., & Hartnoll, S. A. Phonons, electrons and thermal transport in Planckian high $T_c$ materials. *npj Quantum Mater*. **6**, 81 (2021).

[35] Henn, R., Kircher, J., *et al*. Far-infrared *c*-axis response of $La_{1.87}Sr_{0.13}CuO_4$ determined by ellipsometry. *Phys. Rev. B* **53**, 9353 (1996).

[36] Bilbro, L., Aguilar, R., Logvenov, G. *et al.* Temporal correlations of superconductivity above the transition temperature in $La_{2-x}Sr_xCuO_4$ probed by terahertz spectroscopy. *Nature Phys* **7**, 298–302 (2011).